\documentclass[twocolumn,aps,prl,floatfix,superscriptaddress]{revtex4-1}
\usepackage{amsmath,amssymb}
\usepackage{graphicx,float}
\usepackage{times,txfonts}
\usepackage{color}
\usepackage{multirow}
\usepackage{bbold}
\usepackage{color}
\usepackage{soul}
\usepackage{hyperref}
\usepackage{times,txfonts}
\usepackage{natbib}
\usepackage{blindtext}
\usepackage[export]{adjustbox}
\usepackage{mathrsfs}
\usepackage{textcomp}
\usepackage{blkarray}
\usepackage{multirow}

\renewcommand{\epsilon}{\varepsilon}

\def\VR{\kern-\arraycolsep\strut\vrule &\kern-\arraycolsep}
\def\vr{\kern-\arraycolsep & \kern-\arraycolsep}
\usepackage{sistyle}
\usepackage{soul}
\definecolor{lightblue}{RGB}{185,210,248}
\sethlcolor{lightblue}

\begin{document}
\title{Measuring azimuthal and radial modes of photons}

\author{Fr\'ed\'eric Bouchard}
\email{fbouc052@uottawa.ca}
\affiliation{Department of Physics, University of Ottawa, 25 Templeton street, Ottawa, Ontario, K1N 6N5 Canada}
\author{Natalia Herrera Valencia}
\affiliation{Institute for Quantum Optics and Quantum Information (IQOQI), Austrian Academy of Sciences, Boltzmanngasse 3, A-1090 Vienna, Austria}
\affiliation{Universit\'e d\'Aix-Marseille, Centre de Saint-J\'er\^ome, 13014, Marseille, France.}
\author{Florian Brandt}
\affiliation{Institute for Quantum Optics and Quantum Information (IQOQI), Austrian Academy of Sciences, Boltzmanngasse 3, A-1090 Vienna, Austria}
\author{Robert Fickler}
\affiliation{Institute for Quantum Optics and Quantum Information (IQOQI), Austrian Academy of Sciences, Boltzmanngasse 3, A-1090 Vienna, Austria}
\author{Marcus Huber}
\affiliation{Institute for Quantum Optics and Quantum Information (IQOQI), Austrian Academy of Sciences, Boltzmanngasse 3, A-1090 Vienna, Austria}
\author{Mehul Malik}
\affiliation{Institute for Quantum Optics and Quantum Information (IQOQI), Austrian Academy of Sciences, Boltzmanngasse 3, A-1090 Vienna, Austria}
\affiliation{Institute of Photonics and Quantum Sciences (IPaQS), Heriot-Watt University, Edinburgh, EH14 4AS, UK}
%



\begin{abstract}
With the emergence of the field of quantum communications, the appropriate choice of photonic degrees of freedom used for encoding information is of paramount importance. Highly precise techniques for measuring the polarisation, frequency, and arrival time of a photon have been developed. However, the transverse spatial degree of freedom still lacks a measurement scheme that allows the reconstruction of its full transverse structure with a simple implementation and a high level of accuracy. Here we show a method to measure the azimuthal and radial modes of Laguerre-Gaussian beams with a greater than 99~\% accuracy, using a single phase screen. We compare our technique with previous commonly used methods and demonstrate the significant improvements it presents for quantum key distribution and state tomography of high-dimensional quantum states of light. Moreover, our technique can be readily extended to any arbitrary family of spatial modes, such as mutually unbiased bases, Hermite-Gauss, and Ince-Gauss. Our scheme will significantly enhance existing quantum and classical communication protocols that use the spatial structure of light, as well as enable fundamental experiments on spatial-mode entanglement to reach their full potential.
\end{abstract}

\maketitle


%

\section{Introduction}

Photons have long been a candidate of choice for many proof-of-principle experiments in quantum communications, quantum information processing, and foundations of quantum mechanics~\cite{zeilinger:05}. For several decades, due to its maturity in generation and detection, the polarization of photons has now become a standard experimental resource for fundamental and applied experiments, and, to this day, is still being used in a wide array of important experimental demonstrations~\cite{giustina:15,shalm:15,yin:17}. Nevertheless, other photonic degrees of freedom, such as frequency and time~\cite{franson:89,jha:08} or position and momentum~\cite{walborn:07}, offer new, yet unexploited advantages where polarization encoding is limited. One advantage of these degrees of freedom is their high-dimensional nature, whereas polarization is inherently bidimensional, i.e. \emph{qubits}. High-dimensional quantum systems, also known as \emph{qudits}, are both interesting at the fundamental level and useful in applications such as quantum communications and cryptography~\cite{cerf:02,erhard:17}, offering an increased information capacity and greater resistance to noise~\cite{krenn:17}.

A specific family of spatial modes that has gained a lot of attention in the last years is the azimuthal modes of Laguerre-Gauss (LG) beams. In particular, it was found that such solutions of the paraxial wave equation are related to an orbital angular momentum (OAM) proportional to the azimuthal mode index $\ell$~\cite{allen:92} and are characterized by a twisted helical wavefront of the form $\exp \left( i \ell \phi \right)$, where $\phi$ is the azimuthal coordinate. Mathematically, these solutions have the advantage of representing a set of complete and orthogonal functions, thus forming a convenient basis to expand any arbitrary azimuthal functions. Experimentally, OAM modes have been a fruitful testbed for experimental demonstrations in quantum entanglement~\cite{krenn:14,malik:16}, quantum simulation~\cite{cardano:15} and QKD~\cite{vallone:14,mirhosseini:15,sit:17}. Due to their simple form, OAM states are readily generated using devices that shape the wavefront of an incoming beam. Spiral phase plates~\cite{beijersbergen:94}, pitch-fork gratings~\cite{bazhenov:92}, spatial light modulators (SLM)~\cite{heckenberg:92}, and $q$-plates~\cite{marrucci:06} are examples of established devices to generate light beams carrying OAM. Nevertheless, none of the aforementioned techniques directly generate pure LG modes, due to difficulties in manipulating their radial component. Therefore, several generation techniques have been proposed and implemented for experimentally achieving pure LG modes. Among these, holographic amplitude-masking techniques modulating the amplitude and phase of a beam using a single phase-only SLM, although lossy, have been a useful experimental tool to generate any arbitrary desired modes with high precision~\cite{kirk:71,bolduc:13}. For instance, the amplitude modulated holograms may be straightforwardly displayed on an SLM, thus making the generation of LG modes simple and compact for table-top experiments.

Although the generation of spatial modes is relatively simple to realize, it is surprisingly not the case for their detection. A technique known as \emph{phase-flattening} has been demonstrated and become the standard for measuring OAM states of light~\cite{mair:01}. In this scheme, the incoming OAM beam that is to be measured is impinged onto a phase element and subsequently coupled to a single-mode fibre (SMF). For the detection of a given OAM value, a phase pattern with the opposite OAM is imprinted on the SLM, thereby flattening the phase of the incoming beam and allowing it to couple efficiently to the SMF. In all other cases, the resulting beam after the phase modulation will not match the fundamental mode of the SMF. In this way, the phase-flattening method can be used as a filter to measure the OAM content of an unknown incoming beam. This technique also possesses limitations where mode-dependent losses renders the detection of higher-order OAM states less efficient~\cite{qassim:14}. Nevertheless, phase-flattening has now become a standard method for measuring OAM and is also widely used in experiments utilizing OAM both in the classical and the quantum regime.
In another approach, it is possible in principle to measure any arbitrary spatial modes with high accuracy using a large enough sequence of phase elements and free-space propagation separating the elements. For example, with this configuration, a mode sorter was realized using two phase elements~\cite{berkhout:10,mirhosseini:13}, namely an unwrapper and a phase corrector. A sorter-type scheme has the advantage of being able to measure the OAM value of the incoming beam in a single shot, rendering the detection scheme more efficient. Moreover, when considering a larger number of phase elements, a larger flexibility allows one to perform a measurement over a much larger number of modes~\cite{fontaine:18} or to carry out mode transformations~\cite{morizur:10}. However, as the number of phase elements increases, small imperfections in alignment leads to significant mode cross-talks and limits the practicality of such implementations.

Here, we propose and experimentally demonstrate a method that we refer to as \emph{intensity-flattening}. Our method extends the well-established phase-flattening method and enables us to measure arbitrary spatial modes using a simple experimental configuration that requires a single hologram. With a reasonable amount of loss, our method enables us to measure spatial modes with extremely small crosstalk values corresponding to a visibility larger than 99~\%. This technique will be useful for fundamental, proof-of-concept experiments and all tasks where high-quality measurements are necessary and losses can be tolerated.

\section{Theory}

Let us set the stage by writing out the orthogonality condition for the Laguerre-Gauss functions, i.e.

\begin{eqnarray}
\int_{0}^{2 \pi} \int_{0}^{\ \infty} \mathrm{LG}^*_{\ell',p'} (r ,\varphi) \, \mathrm{LG}_{\ell,p} (r,\varphi) \, \mathrm{d}^2 r \, = \delta_{\ell \ell'} \, \delta_{p p'}.
\label{eq:LGorth}
\end{eqnarray}

\begin{figure}[t]
	\begin{center}
	\includegraphics[width=0.95\columnwidth]{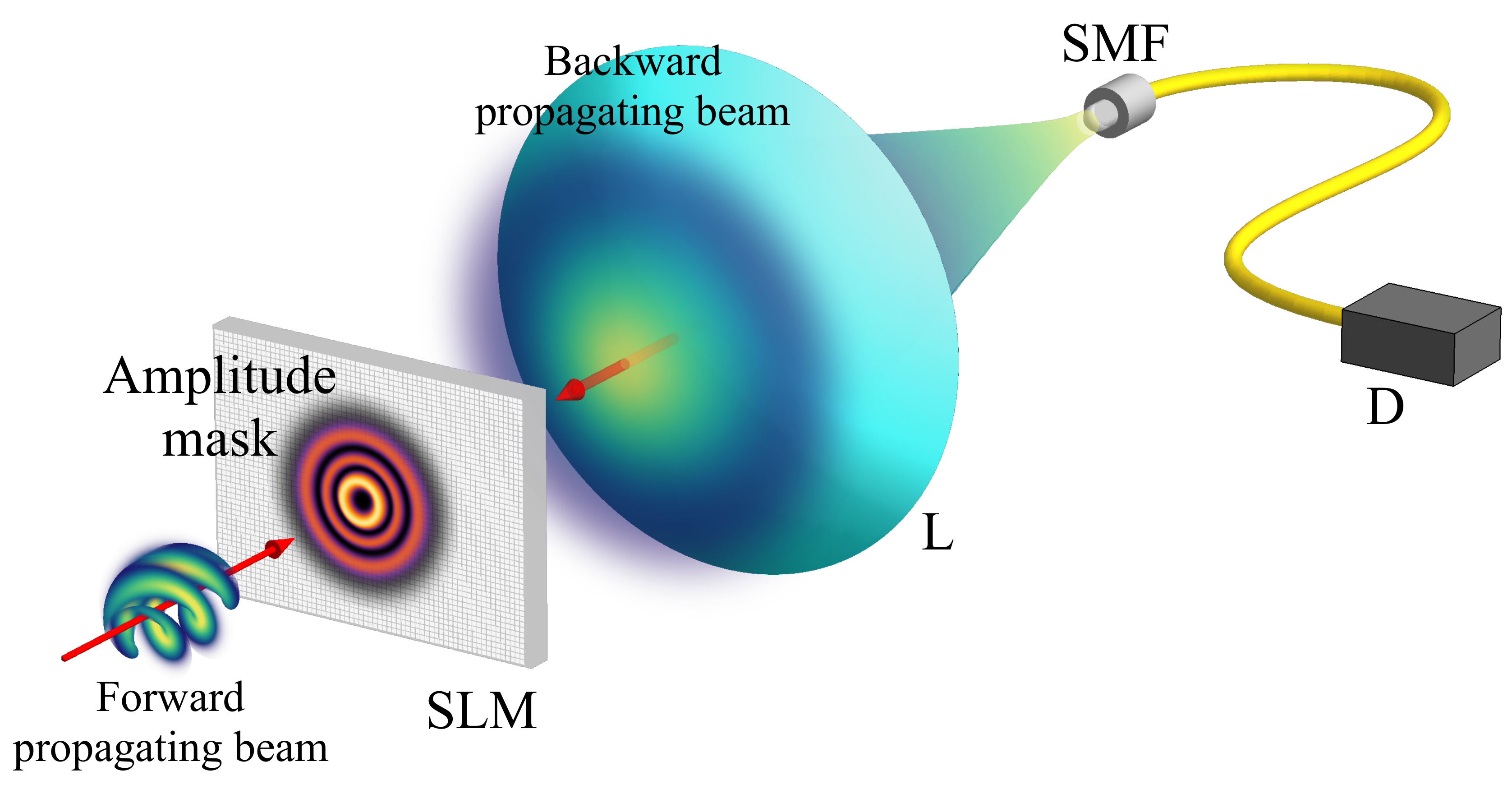}
	\caption[]{\textbf{Simplified experimental setup.} A forward-propagating beam with an unknown spatial mode is made incident on a spatial light modulator (SLM). A hologram simultaneously modulating the phase and the amplitude of the incoming beam is displayed on the SLM. Subsequently, the outgoing beam is coupled to a single-mode fibre (SMF) after passing through a set of lenses and microscope objectives (not shown). The choice of lenses can be understood by considering a back-propagating beam exiting the SMF and made incident on the SLM from the back. According to the intensity-flattening technique presented here, this beam should be expanded on the SLM in order to flatten the intensity distribution of the Gaussian component.}
	\label{fig:setup}
	\end{center}
\end{figure}

In the laboratory, an input LG beam, $\mathrm{LG}_{\ell,p} (r,\varphi)$, is generated using the previously mentioned amplitude-masking method on a first SLM. The beam is now made incident on a second SLM which is displaying the mode $ \mathrm{LG}^*_{\ell',p'} (r, \varphi)$, also using the amplitude-masking technique. Finally, the beam is made to couple to an SMF and the coupled intensity or photon counts are recorded. Due to the unitarity of free-space propagation, we have the freedom to choose at which point we calculate the overlap integral after the SLMs. Thus we imagine a backward propagating beam from the SMF to the second SLM. This experimental scenario corresponds to the overlap integral,

\begin{eqnarray}
\int_{0}^{2 \pi} \int_{0}^{\ r_\mathrm{max}} \mathrm{LG}^*_{\ell',p'} (r, \phi) \, \mathrm{LG}_{\ell,p} (r, \varphi) \, e^{ - r^2/ w_0^2} \, \mathrm{d}^2 r \, \neq \delta_{\ell \ell'} \, \delta_{p p'},
\label{eq:LGgauss}
\end{eqnarray}

where $r_\mathrm{max}$ takes into account finite numerical apertures in the experiment and $w_0$ is the beam waist of the SMF. We note that this integral is different from the overlap integral of LG modes, Eq.~\ref{eq:LGorth}, due to the additional gaussian factor of the SMF~\cite{roux:14,zhang:14}. The aim of our method is to remove the effect of this Gaussian factor to retrieve the standard orthogonality relation of LG beams. A simple but very effective way of achieving this, is by increasing the value of the backward propagating beam waist, $w_0$, in Eq.~\ref{eq:LGgauss}. For a large enough $w_0$, from the perspective of the LG terms in the integral, the additional Gaussian factor will appear flat over the region of interest, thus retrieving Eq.~\ref{eq:LGorth}. This intensity-flattening technique requires minimal modifications to standard experimental setups measuring optical spatial modes, and allows one to select the appropriate trade-off between mode visibility and losses by tuning the beam waist, see Fig.~\ref{fig:setup}. In order to demonstrate this powerful idea experimentally, we build a simple experimental setup allowing us to test our intensity-flattening method in several scenarios -- measuring radial modes, key rates in QKD, and quantum state tomography. We also investigate and compare the performance of the intensity-flattening technique when considering beams other than Gaussian, such as flat-top and exponential, which can be seen in the Supplementary Material. 

\section{Experimental setup}

An attenuated diode laser at a wavelength of 810~nm is coupled to an SMF to clean its spatial profile to the fundamental Gaussian mode. The beam is coupled out of the SMF using a collimator resulting in a beam with a $1/e^2$ beam waist of 1.1~mm, which is then enlarged using a telescope with a magnification of $f_2/f_1 = (300~\mathrm{mm})/(50~\mathrm{mm}) = 6$, where $f_1$ and $f_2$ are the focal length of the first and the second lens in the telescope, respectively. The large collimated beam is made incident on SLM-A where the desired spatial mode is generated using an amplitude-masking technique~\cite{bolduc:13}. The beam waist of the mode displayed on SLM-A is given by $w_0=500~\mathrm{\mu m}$. A $4f$-system is then used in order to filter out the first order of diffraction and to image SLM-A onto SLM-B. The beam is then sent through a second telescope with a magnification of $f_4/f_3 = (50~\mathrm{mm})/(200~\mathrm{mm}) = 0.25$ and then coupled to an SMF using a 10-X microscope objective. The choice of $f_3$ and $f_4$ becomes clearer when considering the backward-propagating beam, (Fig.~\ref{fig:setup}), where the effect of the telescope is to enlarge the size of the backward-propagating beam on SLM-B to 4.2~mm, hence increasing the beam waist of the Gaussian factor in Eq.~\ref{eq:LGgauss}. A more detailed experimental setup can be found in the Supplementary Material.

\begin{figure}[t]
	\begin{center}
	\includegraphics[width=1\columnwidth]{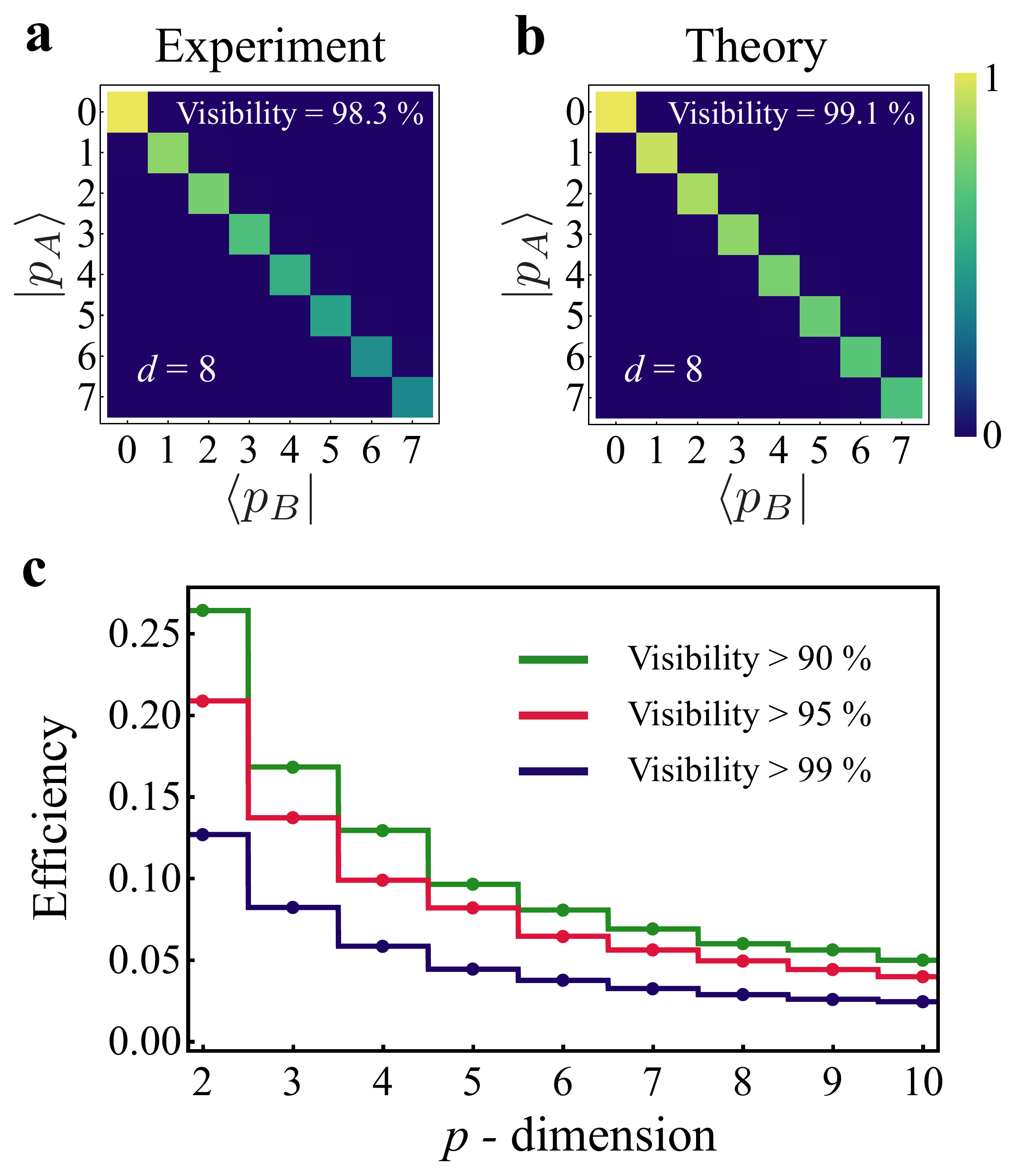}
	\caption[]{\textbf{Measurement of radial modes.} \textbf{a} Experimentally measured and \textbf{b} simulated cross-talk matrix of radial modes ranging from $p=0$ to $p=7$ in a prepare-and-measure setting. The cross-talk matrix is normalized to unity by dividing each elements by the element with maximum counts. The rows and the columns correspond to the states, $| p_A \rangle$ and $| p_B \rangle$, prepared and measured by Alice and Bob, respectively. A visibility of $V=98.3~\%$ is obtained from the experimentally measured cross-talk matrix. In theory, a visibility in excess of 99~\% is achieved by considering a back-propagating with a beam waist 5.4 times larger than that of the beam waist of the detection mode of the holograms. \textbf{c} The efficiency of the intensity-flattening measurement technique is shown as a function of dimensionality of radial modes. For a dimension of $d$, radial modes ranging from $p=0$ to $p=d-1$ are considered. For each dimensions, the reported efficiencies are obtained by increasing the beam waist of the back-propagating up to the point where visibilities are in excess of 99~\% (dark blue), 95~\% (red) and 90~\% (green).}
	\label{fig:radial}
	\end{center}
\end{figure}

\section{Radial modes}

As a first experimental demonstration of our technique, let us consider the radial modes of the LG beams. These modes have recently been investigated both theoretically and experimentally in the context of quantum information~\cite{karimi:14a,karimi:14b,zhang:14,plick:15} and play a key role in fully utlizing the information-carrying capacity of a photon. Since then, several experimental techniques have been proposed to measure radial modes in a sorter configuration, i.e. using a scattering medium~\cite{fickler:17} or taking advantage of the $p$-dependent Gouy phase in an interferometric configuration~\cite{zhou:17,gu:18}. Such schemes have the advantage of having a higher detection efficiency in principle compared to a filter-type measurement as we propose. Nevertheless, in the first case, low transmission efficiencies prohibit its use in a realistic quantum experiment and in the second case, the stability and interferometric nature of the implementation makes these techniques challenging. In contrast, our method has the advantage of being simple, compact and stable for measuring radial modes. In order to demonstrate the quality of measurements achievable with our method, we measure the cross-talk among radial modes ranging from $p=0$ to 7 using our intensity-flattening technique, see Fig.~\ref{fig:radial}. The modal cross-talk is characterized by considering the visibility of the cross-talk matrix, which we define as $V= \sum_i C_{ii} / \sum_{ij} C_{ij}$, where $C_{ij}$ corresponds to the cross-talk matrix. For an 8-dimensional radial mode subspace, we experimentally obtain a visibility value of $V = 98.3~\%$, which is the highest experimentally achieved value so far reported (to the best of our knowledge). As a comparison, the measurement of 8 dimensions of radial modes using a phase-flattening-only scheme achieves a visibility of $V=46.6~\%$, while an amplitude-masking-only scheme achieves 51.0~\%, see Supplementary Material.

We note that we may observe a mode-dependent efficiency in our measurements, which is attributed to the overall transmission of the amplitude mask due to the geometry of the imprinted modes, as well as the coupling to the SMF. 
This effect is also seen from the theory, see Fig.~\ref{fig:radial}-\textbf{b}, and can be straightforwardly compensated for. In order to achieve a visibility of $98.3~\%$, the beam waist of the virtually backward propagating beam has been chosen to be 8.4 times larger than the beam waist of the generation and measurement holograms. We experimentally measured the average efficiency of detection to be 3.2~\%, for radial modes ranging from $p=0$ to 7, where losses due to the amplitude mask and coupling to the single mode fibre are taken into account. In theory, for an 8-dimensional radial mode subspace, a visibility in excess of 99~\% is achieved by enlarging the backward-propagating beam by a factor of 5.4.

By varying the size of the back-propagating beam on SLM-B we may achieve, in theory, arbitrarily high visibility values at the cost of an increase in loss. However, we demonstrate that high visibility may still be achieved with reasonable losses, rendering this technique useful for a broad range of experiments. In general, when considering higher-order modes and thus larger dimensional states, enlarged beam waists $w_0$ must be considered for a similar visibility value. In order to show this effect, we calculate the detection efficiency resulting from increasing the beam waist of the back-propagating beam for obtaining visibility values that are larger than 90~\%, 95~\% and 99~\%, for several dimensions of radial subspaces, see Fig.~\ref{fig:radial}-\textbf{c}. We note that in the case of a 10-dimensional subspace, i.e. $p=0$ to 9, a visibility larger than 99~\% is achieved with an efficiency of 2.5~\%, which is often tolerable in quantum information processing as well as classical application tasks. We further note that our technique has the advantage of allowing the user to vary $w_0$ at will in order to obtain a certain visibility for a tolerable efficiency. 

\section{Full-field modes}

As a second demonstration of our intensity-flattening method for measuring spatial modes, we consider the full-field structure of spatial modes, i.e. the joint azimuthal and radial degrees of freedom. The transverse spatial degree of freedom inherently requires two spatial coordinates to characterize the transverse plane, e.g. $x$ and $y$ in cartesian coordinates, or $r$ and $\phi$ in polar coordinates. Thus in order to take full advantage of transverse spatial modes, it is becoming increasingly important to take into consideration both the azimuthal and radial modes when dealing with LG beams~\cite{zhao:15,kahn:16}. However, due to the lack of a proper technique to measure arbitrary spatial modes in a feasible experimental implementation, only a few experiments have investigated azimuthal and radial modes jointly for quantum entanglement~\cite{salakhutdinov:12,krenn:14} and for classical communications ~\cite{trichili:16}. We now demonstrate how our intensity flattening can be applied in a full-field experiment by measuring states of light in both azimuthal and radial modes. 

\begin{figure}[t]
	\begin{center}
	\includegraphics[width=0.9\columnwidth]{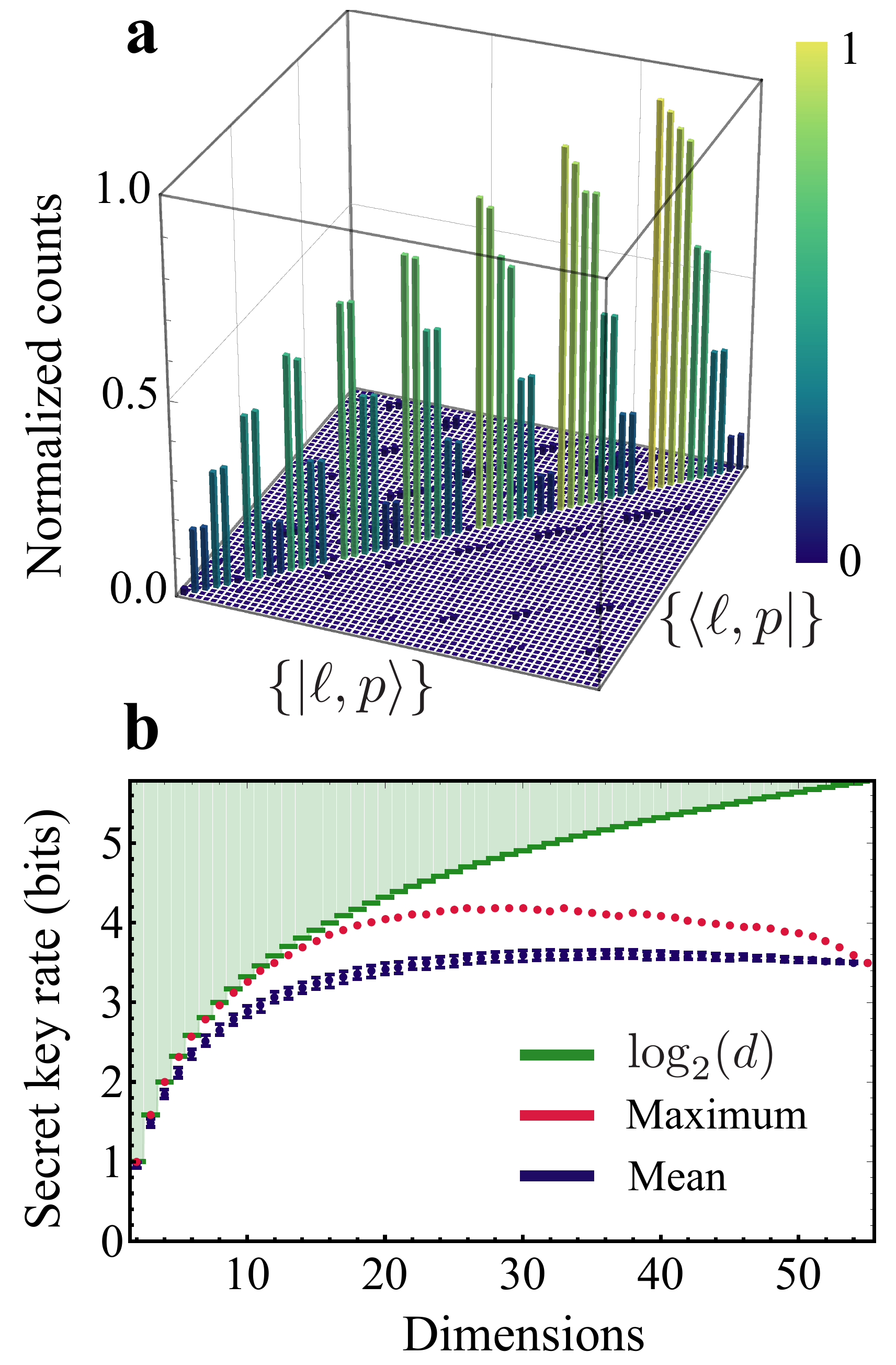}
	\caption[]{\textbf{Measurement of a 55-dimensional space using azimuthal and radial modes.} \textbf{a} Experimentally measured cross-talk matrix for a 55-dimensional space of azimuthal and radial modes. In order to show mode-dependent efficiencies, the cross-talk matrix is normalized to unity by dividing each elements by the element with maximum counts. The list of states $\{ | \ell,p \rangle \}$ are given explicitly in the Supplementary Material. \textbf{b} Secret key rates obtained from lower dimensional subspaces of the full 55-dimensional space. For a given subspace, a sample of 1000 different combinations is selected. For each combination of subspaces, a secret key rate is calculated from the experimental data. The mean and the standard deviation of the secret key rates over the 1000 combinations are shown in dark blue. The maximal secret key rates obtained by searching for the optimal subspace using a genetic algorithm are shown in red. The theoretical maximal values are shown in green, given by $\log_2 (d)$. The shaded region corresponds to values of secret key rates inaccessible for the corresponding dimensions.}
	\label{fig:azrad}
	\end{center}
\end{figure}

It has been shown repeatedly that high-dimensional states of light have various applications in quantum information. However, in any experimental implementations, one is rapidly confronted with the trade-off between higher dimensions and obtaining high quality measurements~\cite{krenn:14,bouchard:18c}. Therefore, for a given quantum information protocol, the optimal dimensionality in experiments dealing with spatial mode is rarely the highest achievable dimension. This is commonly due to the fact that higher-order spatial modes typically result in lower measurement quality due to the complexity of the modes, pixel resolutions, or truncation due to a finite numerical aperture. By taking advantage of the mode order given by $N = 2 p + | \ell| +1$, we consider a 55-dimensional space consisting of the 55 lowest order azimuthal and radial LG modes with mode order ranging from $N=1$ to 10, see Supplementary Material for the list of states employed. A visibility of 92.3~\% is experimentally obtained from the full 55-dimensional cross-talk matrix presented in Fig.~\ref{fig:azrad}-\textbf{a}.

\begin{figure*}[t]
	\begin{center}
	\includegraphics[width=1.95\columnwidth]{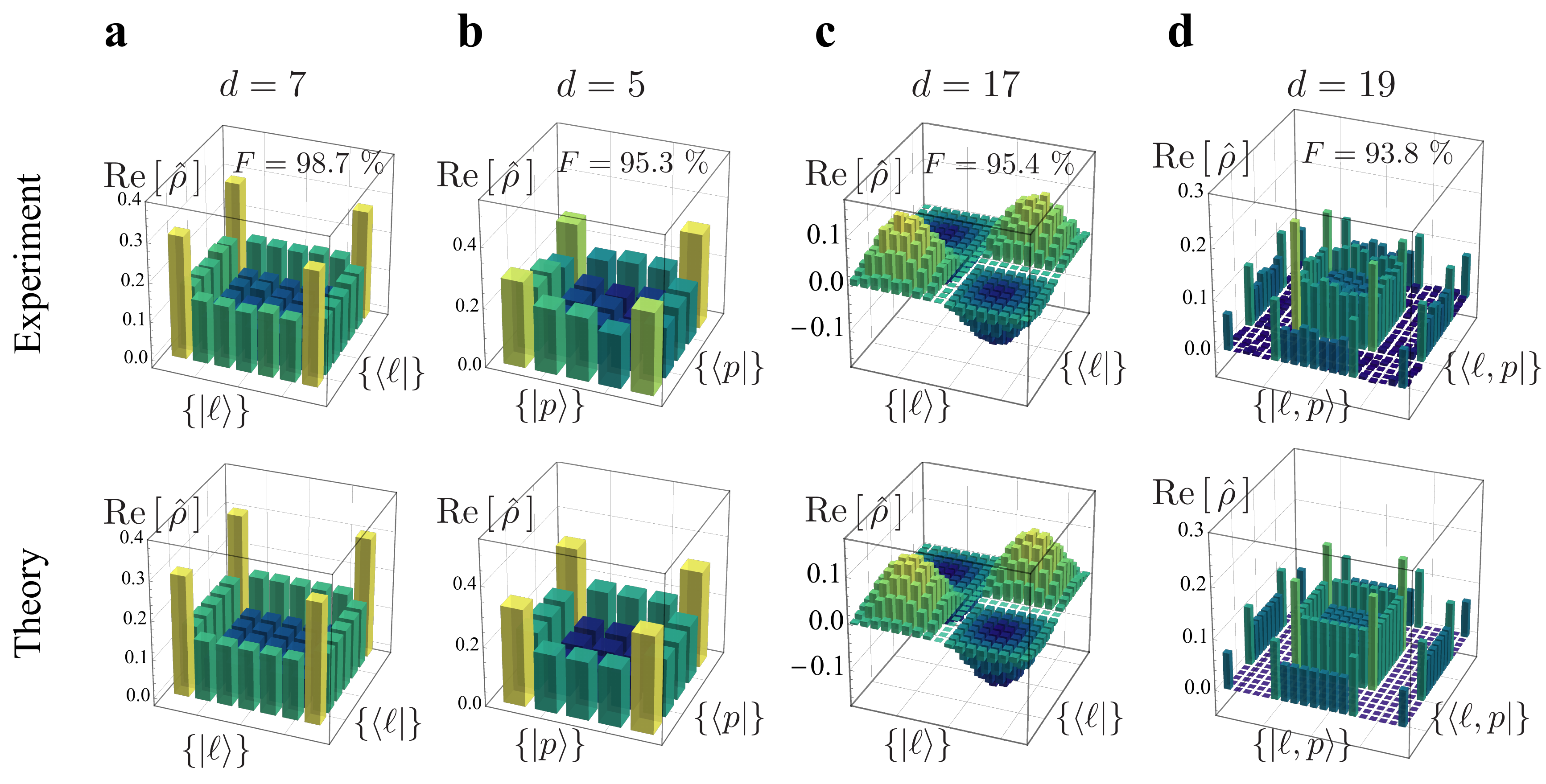}
	\caption[]{\textbf{High-dimensional quantum state tomography.} The experimentally reconstructed density matrices for a \textbf{a} 7-dimensional OAM state, \textbf{b} 5-dimensional radial state, \textbf{c} 17-dimensional OAM state and \textbf{d} 19-dimensional full-field state, are shown in the upper row along with their corresponding theory density matrices, respectively. High-dimensional states giving rise to visually interesting density matrices were chosen in order to resemble \textbf{a}-\textbf{b} a castle, \textbf{c} a sine function, and \textbf{d} a palace, where the explicit forms of the generated states are given in the Supplementary Material. Fidelities of $F=98.7~\%$, 95.3~\%, 95.4~\%, and 93.8~\% were obtained experimentally for \textbf{a}-\textbf{d}, respectively. 
	}
	\label{fig:tomo}
	\end{center}
\end{figure*}

As mentioned previously, there is a trade-off in experiments between dimensionality and visibility. We now explore this interplay by considering lower dimensional subsets of the 55-dimensional cross-talk matrix shown in Fig.~\ref{fig:azrad}-\textbf{a}. An example of a physical parameter that illustrates the trade-off between the dimensionality and the visibility is the secret key rate in a high-dimensional QKD setting. For example, the secret key rate of the high-dimensional BB84 protocol~\cite{cerf:02} is given by $R=\log_2(d)-2 h^{(d)}(e_b)$, where $e_b$ is the quantum bit error rate and $h^{(d)}(x):=-x \log_2 (x/(d-1)) - (1-x) \log_2 (1-x)$ is the $d$-dimensional Shannon entropy. Although we do not perform QKD and only measure in the computational basis, the secret key rate formula provides us with a simple and useful parameter that takes both dimensionality and measurement errors into consideration. For a $d$-dimensional subset of the 55-dimensional space, there are a total of $55! / (55-d)! \, d!$ combinations of possible subspaces. In Fig.~\ref{fig:azrad}-\textbf{b}, we show the average and the standard deviation of the secret key rates obtained from a set of 1000 randomly selected $d$-dimensional subspaces from all possible combinations. Moreover, for a given $d$-dimensional subspace, we search for combination of states that yields the largest secret key rates. However, in some cases, the number of possible combinations becomes extremely large, e.g. for a 27-dimensional subset of the 55-dimensional data, there are a total of $3.8 \times 10^{15}$ possible subsets. Thus, we perform an optimization, consisting of a genetic algorithm, to search among the $d$-dimensional subsets for the optimal secret key rates, see Supplementary Material. The maximal secret key rate is found to be 4.19 bits in a 30-dimensional subspace, corresponding to a visibility of 96.8~\%, which is well above the error bounds for coherent eavesdropping attacks~\cite{cerf:02}. The maximum secret key rates found by the genetic algorithm are, on average, 4 standard deviations larger than the mean values from the random sampling. By doing so, we show another aspect of the potential of high-dimensional states for quantum information protocols by allowing for the careful selection of a lower dimensional subset of the complete data.

\begingroup
\setlength{\tabcolsep}{6pt} 
\renewcommand{\arraystretch}{1.2} 
\begin{table*}
\centering
	\begin{tabular}{lcccc}
	\hline \hline
	Measurements \, \, \,  \, \, \, \, \, \,  & \, \, \, $d$ \, \, \,  & \, \, \, Intensity-flattening \, \, \,  & \, \, \, Phase-flattening \, \, \, & \, \, \, Phase-flattening (AM) \, \, \, \\
		  &   &  (Experiment)  & (Simulation)  &  (Simulation) \\
	\hline 
	Crosstalk - Radial modes & 8 & $V=98.3~\%$ & $V=46.6~\%$ & $V=51.0~\%$  \\
	Secret key rate - Full-field & 30 & $R=4.19~\mathrm{bits} $ & $R=1.34~\mathrm{bits}$ & $R=0.92~\mathrm{bits}$  \\
	QST - Radial modes & 5 & $F=95.3~\%$ & $F=22.3~\%$ & $F=45.0~\%$  \\
	QST - Azimuthal modes & 17 & $F=95.4~\%$ & $F=57.1~\%$ & $F=73.1~\%$  \\
	QST - Full-field modes & 19 & $F=93.8~\%$ & $F=40.9~\%$ & $F=65.2~\%$  \\
	 \hline \hline
	\end{tabular}
\caption[]{\textbf{Comparison of intensity-flattening and phase-flattening for measuring azimuthal and radial modes.} Different sets of measurements are compared in terms of visibility ($V$), secret key rate ($R$), and state fidelity ($F$) for several different dimensions ($d$). The comparison is carried out among intensity-flattening (experimental), phase-flattening (simulated), and phase-flattening with amplitude-masking (simulated).}
\label{table:1}
\end{table*}
\endgroup

\section{High-dimensional state tomography}

As a final test of the versatility and effectiveness of our method, we take on the demanding task of performing high-dimensional quantum state tomography (QST). In particular, we perform our tomographic reconstruction using mutually unbiased bases (MUBs), which are known for dimensions that are power of prime numbers~\cite{durt:10}. The measurements of MUBs is an important task in many high-dimensional quantum information protocols, such as QKD~\cite{mafu:13}, channel characterization~\cite{bouchard:18:d}, and high-dimensional entanglement certification~\cite{bavaresco:18}. 

We start by performing QST of a 7-dimensional OAM space. As a non-trivial state to produce in the laboratory, we consider states that are visually interesting, see Fig.~\ref{fig:tomo}. In order to avoid systematic errors in our tomographic reconstruction~\cite{schwemmer:15}, we experimentally reconstruct the density matrix using a direct inversion given by $\hat{\rho} = \sum_{\alpha,m} P^{(\alpha)}_m \Pi^{(\alpha)}_m - \hat{\mathbb{1}}$, where $\alpha$ labels the MUB, $m$ labels the state, $P^{(\alpha)}_m$ corresponds to the probability of measuring the state $| \psi_m^{(\alpha)} \rangle$ from the MUB $\alpha$ and $\Pi^{(\alpha)}_m$ corresponds to the projector $| \psi_m^{(\alpha)} \rangle \langle \psi_m^{(\alpha)}|$. The experimental generation and reconstruction may be evaluated using the state fidelity given by $F=\left( \mathrm{Tr} \sqrt{\sqrt{\hat{\rho}} \,\, \hat{\rho}_\mathrm{th} \sqrt{\hat{\rho}} }\right)^2$, which reduces to $F= \langle \psi_\mathrm{th} | \, \hat{\rho} \, | \psi_\mathrm{th} \rangle$ in our case since $\hat{\rho}_\mathrm{th}$ are pure states. The experimentally reconstructed 7-dimensional OAM state is shown in Fig.~\ref{fig:tomo}-\textbf{a}, along with its theoretical counterpart. The state fidelity is given $F=98.7~\%$, which shows the high measurement quality of our method for OAM states.

While several techniques have been proposed to measure radial modes, there has been no experimental demonstration of measurements of MUBs for radial modes. To demonstrate the extent of the capability of the intensity-flattening technique, we perform QST of a 5-dimensional state consisting of radial modes ranging from $p=0$ to $p=4$ using MUBs. The experimentally reconstructed density matrix, with a corresponding state fidelity of $F=95.3~\%$, is shown in Fig.~\ref{fig:tomo}-\textbf{b}. We note that the highest radial mode, i.e. $p=4$, corresponds to a mode order of $N_\mathrm{max}=9$. As a comparison, we also perform QST on a 17-dimensional OAM state, where the highest OAM value is $|\ell|=8$, corresponding to a maximal mode order of $N_\mathrm{max} = 9 $ as well, see Fig.~\ref{fig:tomo}-\textbf{c}. Even in such large dimensions, a relatively high state fidelity of 95.4~\% is achieved. Finally, we perform QST on a 19-dimensional state of combined azimuthal and radial modes, see Fig.~\ref{fig:tomo}-\textbf{d}. A state fidelity of $F=93.8~\%$ is obtained from the full-field QST showing the full experimental power of our technique. Table~\ref{table:1} summarizes how our intensity-flattening technique compares with the well-established phase-flattening methods (with and without amplitude masking), when applied to radial mode crosstalk, QKD key rates achievable, and QST of high-dimensional states (details in the Supplementary Material). As can be seen, the intensity-flattening method enables significant improvements on all fronts, including a four-fold increase in key rates and vastly better tomographic state fidelities.

\section{Conclusion}

In conclusion, we have proposed and experimentally demonstrated an intensity-flattening technique which enabled us to projectively measure arbitrary spatial modes with a high level of accuracy. Our method uses a single phase screen, and has the advantage of being simple and straightforward to implement, making it a powerful experimental tool for quantum and classical experiments with the spatial modes of light. In order to demonstrate the versatility of our technique, we have measured radial modes with higher visibilities than ever before -- 98.3~\% in an 8-dimensional state space. Moreover, we have characterized extremely large high-dimensional states with high visibility by combining the azimuthal and radial modes of LG beams, thus taking advantage of the full information capacity of transverse spatial modes. Finally, as an ultimate test of the generality of this technique, we have performed quantum state tomography on high-dimensional azimuthal and radial modes with significant improvements in fidelity over previous measurement techniques. By enabling the precise measurement of the azimuthal and radial modes of light, our method opens a pathway towards practical quantum and classical communication protocols with record information capacities and levels of security.

\bibliographystyle{naturemag}

\vspace{0.2cm}
\noindent 
\vspace{0.5 EM}

\noindent\textbf{Acknowledgments} We thank A. Zeilinger and E. Karimi for many fruitful discussions. F.B. acknowledges the support of the Vanier Canada Graduate Scholarships Program and the Natural Sciences and Engineering Research Council of Canada (NSERC) Canada Graduate Scholarships program. M.M., R.F. and M.H. acknowledge funding from the Austrian Science Fund (FWF) through the START project Y879-N27 and the joint Czech-Austrian project MultiQUEST (I 3053-N27 and GF17-33780L). M.M. acknowledges support from the QuantERA ERA-NET Co-fund (FWF project I 3553-N36) and the Engineering and Physical Sciences Research Council (EPSRC) (EP/P024114/1).

\vspace{0.5 EM}


\vspace{0.5 EM}


\end{document}